\documentclass[twocolumn,floats,showpacs,superscriptaddress,pre]{revtex4}
\usepackage{amsmath,amssymb,bm}
\usepackage{graphics}
\usepackage{epsfig}
\usepackage{graphicx}

\begin{document}

\title{Generalized Canonical Form of a Multi-Time Dynamical Theory and
Quantization}
\author{Natalia Gorobey, Alexander Lukyanenko}
\email{alex.lukyan@rambler.ru}
\affiliation{Department of Experimental Physics, St. Petersburg State Polytechnical
University, Polytekhnicheskaya 29, 195251, St. Petersburg, Russia}
\author{Inna Lukyanenko}
\email{inna.lukyanen@gmail.com}
\affiliation{Institut f\"{u}r Mathematik, TU Berlin, Strasse des 17 Juni 136, 10623
Berlin, Germany}

\begin{abstract}
A generalized canonical form of multi-time dynamical theories is proposed.
This form is a starting point for a modified canonical quantization
procedure of theories based on a quantum version of the action principle. As
an example, the Fokker theory of a direct electromagnetic interaction of
charges is considered.
\end{abstract}

\maketitle
\date{\today }





\section{\textbf{INTRODUCTION}}

The well-known standard canonical quantization procedure is formulated for
ordinary dynamical theories with a single parameter of time $t$, and\ the
action of which may be written in a Lagrangian form: 
\begin{equation}
I=\int L\left( q,\overset{\cdot }{q}\right) dt.  \label{1}
\end{equation}%
where $L\left( q,\overset{\cdot }{q}\right) $ is a Lagrangian function, the
dot denotes the derivative with respect to the time parameter $t$. This
procedure of canonical quantization originates from a canonical form of the
classical theory. 
A canonical momentum for a dynamical variable $q_{i},i=1,2,...,n,$, where $n$
is the dimensionality of a configuration space of a system, is defined as
follows:
\begin{equation}
p_{i}=\frac{\partial L\left( q,\overset{\cdot }{q}\right) }{\partial \overset%
{\cdot }{q}_{i}},  \label{2}
\end{equation}
and then the Hamiltonian is defined by the Legendre transformation: 
\begin{equation}
H\left( p,q\right) \equiv \left. \left( \sum_{i=1}^{n}p_{i}\overset{\cdot }{q%
}_{i}-L\left( q,\overset{\cdot }{q}\right) \right) \right\vert _{\overset{%
\cdot }{q}=\overset{\cdot }{q}\left( p,q\right) },  \label{3}
\end{equation}%
where velocities are supposed to be excluded by use of Eq. (\ref{2}). We do
not consider here singular theories with constraints. This will be a subject
of a subsequent work. The quantization (in a coordinate representation) is
performed by the replacement of canonical variables $\left(
q_{i},p_{i}\right) $ by operators acting on a wave function $\psi \left(
q,t\right) $: 
\begin{eqnarray}
\widehat{q}_{k}\psi &\equiv &q_{k}\psi ,  \notag \\
\widehat{p}_{k}\psi &\equiv &\frac{\hslash }{i}\frac{\partial \psi }{%
\partial q_{k}}.  \label{4}
\end{eqnarray}

However, a quantization of a multi-time dynamical theory as the Fokker
theory of direct electromagnetic interaction of charges \cite{F}, has a
practical interest. In this theory a point charge $e_{a}$ has its own time
parameter $t_{a}$, and the action can not be written in the Lagrangian form (%
\ref{1}). In the preprint series \cite{GL,GL1} a new form of quantum
mechanics in terms of a quantum action principle was proposed. It was noted
that the new approach is the most appropriate for multi-time dynamical
theories. To perform the modified canonical quantization procedure in that
case, one needs a generalized canonical form of the action. In the present
work such a form of the action for a multi-time dynamical theory is
proposed. As an example, the Fokker theory of a direct electromagnetic
interaction of two charges is considered.

\section{MULTI-TIME DYNAMICAL THEORY}

Let the dynamics of a system be described by a several number of sets of
dynamical variables $q_{a}\left( t_{a}\right) $ with an own for each set
time parameter $t_{a}\in \left[ 0.T_{a}\right] $. Numerating indices of
dynamical variables in each set are omitted for brevity. We consider each
set of dynamical variables $q_{a}$ as coordinates of a particle in a
configuration space. Let the action of that system be a smooth functional of
coordinates of particles and their velocities: 
\begin{equation}
I=I\left[ q_{a}\left( t_{a}\right) ,\overset{\cdot }{q}_{a}\left(
t_{a}\right) \right] ,  \label{5}
\end{equation}%
where the dot denotes the derivative with respect to the time parameter $%
t_{a}$ corresponding to the particle $a$. An example of such dynamical
system gives the Fokker theory of direct electromagnetic interaction of two\
charges described by the action (the velocity of light is taken equal unity)
\cite{F}: 
\begin{eqnarray}
I_{F} &=&-m_{1}\int ds_{1}-m_{1}\int ds_{2}  \notag \\
&&+\frac{1}{2}e_{1}e_{2}\int dx_{1}^{\mu }\int dx_{2\mu }\delta \left(
s_{12}^{2}\right) .  \label{6}
\end{eqnarray}%
Here $x_{1,2}^{\mu },\mu =0,1,2,3$ are coordinates of charges in the
Minkowsky space, indices are lowered and rised by means of the metric $\eta
_{\mu \nu }\equiv diag\left( +1,-1,-1,-1\right) $, and short notations for
squares and scalar products of vectors are used:
\begin{eqnarray*}
ds_{1,2}^{2} &\equiv &dx_{1,2}^{\mu }dx_{1,2\mu }, \\
s_{1,2}^{2} &\equiv &\left( x_{1}^{\mu }-x_{2}^{\mu }\right) \left( x_{1\mu
}-x_{2\mu }\right) \equiv \left( x_{1}-x_{2}\right) ^{2}.
\end{eqnarray*}%
If we take $x_{1,2}^{0}\equiv t_{1,2}$, and $\left\{
x_{1,2}^{i},i=1,2,3\right\} \equiv q_{1,2}$, the Fokker action (\ref{6}) may
be written in the form (\ref{5}):
\begin{eqnarray}
\!I &=&-m_{1}\int\limits_{0}^{T_{1}}\sqrt{1-\overset{\cdot }{q}_{1}^{2}}%
dt_{1}-m_{2}\int\limits_{0}^{T_{2}}\sqrt{1-\overset{\cdot }{q}_{2}^{2}}dt_{2}
\label{7} \\
&&{+}\frac{1}{2}\int\limits_{0}^{T_{1}}dt_{1}\int\limits_{0}^{T_{2}}dt_{2}%
\delta \left( (t_{1}{-}t_{2})^{2}{-}(q_{1}{-}q_{2})^{2}\right) \left( 1{-}%
\overset{\cdot }{q}_{1}\overset{\cdot }{q}_{2}\right) .  \notag
\end{eqnarray}%
The multi-time dynamics of the system, which follows from the stationarity
condition for the action (\ref{3}), is determined by the system of
Euler-Lagrange (EL) equations for each particle in its own time parameter $%
t_{a}$: 
\begin{equation}
\frac{\delta I}{\delta q_{a}}-\frac{d}{dt_{a}}\frac{\delta I}{\delta \overset%
{\cdot }{q}_{a}}=0  \label{8}
\end{equation}%
In the case of Fokker theory with the action (\ref{6}), the set of equations
(\ref{8}) is reduced to two ($3D$-vector) integro-differential equations.

\section{GENERALIZED CANONICAL FORM OF MULTI-TIME DYNAMICAL THEORY}

Let us define canonical momenta of particles as variational derivatives of
the action (\ref{5}) with respect to velocities: 
\begin{equation}
p_{a}\left( t_{a}\right) \equiv \frac{\delta I}{\delta \overset{\cdot }{q}%
_{a}\left( t_{a}\right) }.  \label{9}
\end{equation}%
If the action has the Lagrangian form (\ref{1}), this definition is
equivalent to the ordinary definition of canonical momenta. Let us assume
that the set of equations (\ref{9}) can be solved with respect to
velocities, and the solution is: 
\begin{equation}
\overset{\cdot }{q}_{a}\left( t_{a}\right) =F_{a}\left( t_{a};\left[
q_{a}\left( t_{a}\right) ,p_{a}\left( t_{a}\right) \right] \right) .
\label{10}
\end{equation}%
This solution is a function of a corresponding time parameter and a
functional of a trajectory $\left( q_{a}\left( t_{a}\right) ,p_{a}\left(
t_{a}\right) \right) $ of the system in its phase space. The following step
in our approach differs from a similar step in the ordinary canonical
theory. Let us define a generalized Hamiltonian by means of generalized
Legendre transformation as follows: 
\begin{eqnarray}
\!\!\!\!\!\!\!\!H\left[ q_{a}\left( t_{a}\right) ,p_{a}\left( t_{a}\right) %
\right]  &\equiv &\left[ \sum\limits_{a}\int\limits_{0}^{T_{a}}dt_{a}p_{a}%
\left( t_{a}\right) \overset{\cdot }{q}_{a}\left( t_{a}\right) \right.
\notag \\
&&\left. \left. -I\left[ q_{a}\left( t_{a}\right) ,\overset{\cdot }{q}%
_{a}\left( t_{a}\right) \right] \right] \right\vert _{\overset{\cdot }{q}=F}.
\label{11}
\end{eqnarray}%
Then the action may be written in a generalized canonical form: 
\begin{eqnarray}
I\left[ q_{a}\left( t_{a}\right) ,p_{a}\left( t_{a}\right) \right]
&=&\sum\limits_{a}\int\limits_{0}^{T_{a}}dt_{a}p_{a}\left( t_{a}\right)
\overset{\cdot }{q}_{a}\left( t_{a}\right)   \label{12} \\
&&-H\left[ q_{a}\left( t_{a}\right) ,p_{a}\left( t_{a}\right) \right] .
\notag
\end{eqnarray}%
Let us check, first of all, that the condition of stationarity of the action
(\ref{12}) with respect to independent variations of coordinates $%
q_{a}\left( t_{a}\right) $\ of particles, and their momenta $p_{a}\left(
t_{a}\right) ,$\ is equivalent to the system of EL equations (\ref{8}). It
is usefull to introduce at this stage a generalized Poisson brackets defined
by the canonical commutational relations (all other brackets are zero): 
\begin{equation}
\left\{ q_{ai}\left( t_{a}\right) ,p_{ak}\left( \widetilde{t}_{a}\right)
\right\} =\delta _{ik}\delta \left( t_{a}-\widetilde{t}_{a}\right) .
\label{13}
\end{equation}%
"Inner" indices of the canonical variables are written here in the open
form. The stationarity condition for the action (\ref{12}) can be written
with the use of these brackets as a system of generalized canonical
relations: 
\begin{equation}
\left\{ q_{a}\left( t_{a}\right) ,I\right\} =\left\{ p_{a}\left(
t_{a}\right) ,I\right\} =0.  \label{14}
\end{equation}%
Here, the first bracket equals is zero according to the following
equalities: 
\begin{eqnarray}
\left\{ q_{a}\left( t_{a}\right) ,I\right\}  &=&\overset{\cdot }{q}%
_{a}\left( t_{a}\right) -\left\{ q_{a}\left( t_{a}\right) ,H\right\} =%
\overset{\cdot }{q}_{a}\left( t_{a}\right)   \notag \\
&&-\left[ \overset{\cdot }{q}_{a}\left( t_{a}\right)
+\sum\limits_{b}\int\limits_{0}^{T_{b}}dt_{b}p_{b}\left( t_{b}\right) \frac{%
\delta F_{b}}{\delta p_{a}\left( t_{a}\right) }\right.   \notag \\
&&\left. -\sum\limits_{b}\int\limits_{0}^{T_{b}}dt_{b}\frac{\delta I}{\delta
\overset{\cdot }{q}_{b}\left( t_{b}\right) }\frac{\delta F_{b}}{\delta
p_{a}\left( t_{a}\right) }\right]   \notag \\
&=&0,  \label{15}
\end{eqnarray}%
where the definition of momenta (\ref{9}) and a solution of these equalities
with respect to velocities (\ref{10}) were used. The second bracket in Eq. (%
\ref{14}) also equals to zero due to the EL equations (\ref{8}): 
\begin{eqnarray}
\left\{ p_{a}\left( t_{a}\right) ,I\right\}  &=&\overset{\cdot }{p}%
_{a}\left( t_{a}\right) -\left\{ p_{a}\left( t_{a}\right) ,H\right\} =%
\overset{\cdot }{p}_{a}\left( t_{a}\right)   \notag \\
&&-\left[ -\sum\limits_{b}\int\limits_{0}^{T_{b}}dt_{b}p_{b}\left(
t_{b}\right) \frac{\delta F_{b}}{\delta q_{a}\left( t_{a}\right) }\right.
\notag \\
&&+\sum\limits_{b}\int\limits_{0}^{T_{b}}dt_{b}\frac{\delta I}{\delta
\overset{\cdot }{q}_{b}\left( t_{b}\right) }\frac{\delta F_{b}}{\delta
q_{a}\left( t_{a}\right) }  \notag \\
&&\left. +\frac{\delta I}{\delta q_{a}\left( t_{a}\right) }\right]   \notag
\\
&=&0.  \label{16}
\end{eqnarray}%
Therefore, the stationarity condition for the generalized canonical form of
the action (\ref{12}) is equivalent to the original equations of motion (\ref%
{8}) of multi-time dynamical system.

\section{GENERALIZED CANONICAL FORM OF THE FOKKER THEORY}

In the Fokker theory, equations (\ref{9}) form the set of integral equations
with respect to velocities: 
\begin{eqnarray}
p_{1i}\left( t_{1}\right) &=&-\frac{m_{1}\overset{\cdot }{q}_{1i}\left(
t_{1}\right) }{\sqrt{1-\overset{\cdot }{q}_{1}^{2}}}  \notag \\
&&-\frac{1}{2}e_{1}e_{2}\int\limits_{0}^{T_{2}}dt_{2}\overset{\cdot }{q}%
_{2i}\left( t_{2}\right) \delta \left( s_{12}^{2}\right) ,  \notag \\
p_{2i}\left( t_{2}\right) &=&-\frac{m_{2}\overset{\cdot }{q}_{2i}\left(
t_{2}\right) }{\sqrt{1-\overset{\cdot }{q}_{2}^{2}}}  \notag \\
&&-\frac{1}{2}e_{1}e_{2}\int\limits_{0}^{T_{1}}dt_{1}\overset{\cdot }{q}%
_{1i}\left( t_{1}\right) \delta \left( s_{12}^{2}\right) .  \label{17}
\end{eqnarray}%
One can solve this set by use of the perturbation theory with respect to the
parameter $e_{1}e_{2}$ of the direct interaction. In the first order in the
parameter $e_{1}e_{2}$ we have: 
\begin{eqnarray}
\overset{\cdot }{q}_{1i}\left( t_{1}\right) &=&-\frac{p_{1i}\left(
t_{1}\right) }{\sqrt{p_{1}^{2}+m_{1}^{2}}}  \notag \\
&&+\frac{1}{2}\int\limits_{0}^{T_{2}}dt_{2}\frac{e_{1}e_{2}\delta \left(
s_{12}^{2}\right) }{\sqrt{p_{1}^{2}+m_{1}^{2}}\sqrt{p_{2}^{2}+m_{2}^{2}}}
\notag \\
&&\times \left[ p_{1i}\frac{p_{1}p_{2}}{p_{1}^{2}+m_{1}^{2}}-p_{2i}\right] ,
\notag \\
\overset{\cdot }{q}_{2i}\left( t_{2}\right) &=&-\frac{p_{2i}\left(
t_{2}\right) }{\sqrt{p_{2}^{2}+m_{2}^{2}}}  \notag \\
&&+\frac{1}{2}\int\limits_{0}^{T_{1}}dt_{1}\frac{e_{1}e_{2}\delta \left(
s_{12}^{2}\right) }{\sqrt{p_{1}^{2}+m_{1}^{2}}\sqrt{p_{2}^{2}+m_{2}^{2}}}
\notag \\
&&\times \left[ p_{2i}\frac{p_{1}p_{2}}{p_{2}^{2}+m_{2}^{2}}-p_{1i}\right] .
\label{18}
\end{eqnarray}%
In the first order in the parameter $e_{1}e_{2}$ the generalized Hamiltonian
defined by Eq. (\ref{11}) is 
\begin{eqnarray}
&&H\left[ p_{1}\left( t_{1}\right) ,q_{1}\left( t_{1}\right) ,p_{2}\left(
t_{2}\right) ,q_{2}\left( t_{2}\right) \right]  \notag \\
&=&\int\limits_{0}^{T_{1}}\sqrt{p_{1}^{2}+m_{1}^{2}}dt_{1}+\int%
\limits_{0}^{T_{2}}\sqrt{p_{2}^{2}+m_{2}^{2}}dt_{2}  \notag \\
&&+\frac{e_{1}e_{2}}{2}\int\limits_{0}^{T_{1}}dt_{1}\int%
\limits_{0}^{T_{2}}dt_{2}\delta \left( s_{12}^{2}\right)  \notag \\
&&\times \left( 1-\frac{p_{1}p_{2}}{\sqrt{p_{1}^{2}+m_{1}^{2}}\sqrt{%
p_{2}^{2}+m_{2}^{2}}}\right)  \label{19}
\end{eqnarray}%
The first two terms in the right hand side of Eq. (\ref{19}) are Hamiltonian
parts of the canonical action of free particles and two parts of the third
term describe, in the first order in $e_{1}e_{2}$, their Coulomb and
magnetic interactions.

\section{QUANTUM ACTION PRINCIPLE}

Using the generalized canonical form of the action of the multi-time
dynamical theory, we can perform a modified canonical quantization procedure
proposed in the work \cite{GL}. The central object in the modified quantum
theory is an action operator $\widehat{I}$ defined in a space of wave
functionals. In turn, a wave functional $\Psi \left[ q_{a}\left(
t_{a}\right) \right]$ is a functional of a trajectory $q_{a}\left(
t_{a}\right) $ of the system with fixed end points in the configuration
space. To define the action operator, we represent the canonical variables
as operators in the space of wave functionals as follows: 
\begin{eqnarray}
\widehat{q}_{ak}\left( t_{a}\right) \Psi &\equiv &q_{ak}\left( t_{a}\right)
\Psi ,  \notag \\
\widehat{p}_{ak}\left( t_{a}\right) \Psi &\equiv &\frac{\widetilde{\hslash }%
}{i}\frac{\delta \Psi }{\delta q_{ak}\left( t_{a}\right) }.  \label{20}
\end{eqnarray}%
Here we write in the open form the "inner" indice $k=1,2,...,n$. The
constant $\widetilde{\hslash }$ differs from the "ordinary" Plank constant $%
\hslash $, in particular, its physical dimensionality is $\left[ \widetilde{%
\hslash }\right] =Joule\cdot s^{2}$. A relationship between two constants
can be established after a determination of observables in the proposed
quantum theory and comparison between the theory and an experiment. The
operators (\ref{20}) are formally Hermitian with respect to the scalar
product in the space of wave functionals: 
\begin{eqnarray}
&&\left( \Psi _{1},\Psi _{2}\right)  \label{21} \\
&=&\int \prod\limits_{a}\prod\limits_{t_{a}}d^{n}q_{a}\left( t_{a}\right)
\overline{\Psi }_{1}\left[ q_{a}\left( t_{a}\right) \right] \Psi _{2}\left[
q_{a}\left( t_{a}\right) \right] .  \notag
\end{eqnarray}%
A wave functional $\Psi \left[ q_{a}\left( t_{a}\right) \right] $ has now
the natural probabilistic interpretation: $\left\vert \Psi \left[
q_{a}\left( t_{a}\right) \right] \right\vert ^{2}$ is a probability density
of that a system moves along a trajectory in a neighbourhood of given
trajectory $q_{a}\left( t_{a}\right) $. The operators (\ref{20}) obey the
following permutation relations (all other commutators are zero): 
\begin{equation}
\left[ \widehat{q}_{ai}\left( t_{a}\right) ,\widehat{p}_{ak}\left(
\widetilde{t}_{a}\right) \right] =i\widetilde{\hslash }\delta _{ik}\delta
\left( t_{a}-\widetilde{t}_{a}\right) .  \label{22}
\end{equation}%
Replacing canonical variables in the action (\ref{12}) by their operator
realization (\ref{18}), one obtains an action operator $\widehat{I}$. This
operator is also formally Hermitian with respect to the scalar product (\ref%
{21}).

Now, let us return to the formulation of the quantum action principle. For
the action operator $\widehat{I}$ we consider the eigenvalue problem: 
\begin{equation}
\widehat{I}\Psi =\Lambda \Psi .  \label{23}
\end{equation}%
In a general case the set of eigenvalues $\Lambda $\ is parametrized by a
set of parameters which form a smooth manifold. The stationarity condition
for this smooth function of free parameters forms the content of the quantum
action principle. To make this formulation more concrete, let us consider
the Fokker theory. In this case the action operator is 
\begin{eqnarray}
\widehat{I} &=&\int\limits_{0}^{T_{1}}dt_{1}\overset{\cdot }{q}_{1k}\left(
t_{1}\right) \frac{\widetilde{\hslash }}{i}\frac{\delta }{\delta
q_{1k}\left( t_{1}\right) }  \notag \\
&&+\int\limits_{0}^{T_{2}}dt_{2}\overset{\cdot }{q}_{2k}\left( t_{2}\right)
\frac{\widetilde{\hslash }}{i}\frac{\delta }{\delta q_{2k}\left(
t_{2}\right) }  \notag \\
&&-\widehat{H},  \label{24}
\end{eqnarray}%
where the operator $\widehat{H}$ is obtained by the substitution of the
operators (\ref{20}) into the generalized Hamiltonian of the Fokker theory.
In the first order in $e_{1}e_{2}$, the latter one is given by the
expression (\ref{21}). Even in this approximation this expression is too
complex due to the presence of square routs, in particular, in the
denominator of the last term. To simplify the subsequent consideration let
us consider a non-relativistic limit of the theory. In this limit we have:
\begin{eqnarray}
\widehat{H} &=&-\int\limits_{0}^{T_{1}}dt_{1}\frac{\widetilde{\hslash }^{2}}{%
2m_{1}}\frac{\delta ^{2}}{\delta q_{1}^{2}\left( t_{1}\right) }%
-\int\limits_{0}^{T_{2}}dt_{2}\frac{\widetilde{\hslash }^{2}}{2m_{2}}\frac{%
\delta ^{2}}{\delta q_{2}^{2}\left( t_{2}\right) }  \notag \\
&&+\frac{1}{2}e_{1}e_{2}\int\limits_{0}^{T_{1}}dt_{1}\int%
\limits_{0}^{T_{2}}dt_{2}\delta \left( s_{12}^{2}\right) .  \label{25}
\end{eqnarray}%
Here, the last term in the right hand side describes with the relativistic
accuracy the Coulomb interaction of charges. The problem is that this term
is a non-analytic functional of trajectories of charges. In the case of
analytic potentials, a concrete formulation of the action principle was
proposed in the preprint \cite{GL1}. To use this formulation, let us
regularize the Coulomb potential in Eq. (\ref{25}), replacing the $\delta $%
-function by the exponent: 
\begin{equation}
\frac{1}{\sqrt{2\pi \sigma }}\exp \left( -\frac{\left( s_{12}^{2}\right) ^{2}%
}{2\sigma }\right) .  \label{26}
\end{equation}%
At the final stage of calculations the limit $\sigma \rightarrow 0$ is
assumed. After this regularization the interaction term becomes an analytic
functional of trajectories of charges, and the formulation of the quantum
action principle proposed in \cite{GL1} is also applicable in the case under
consideration. Formulation of the quantum action principle in general case
will depend, in particular, on the definition of the square route of the
functional-differential operator: 
\begin{equation}
\sqrt{-\widetilde{\hslash }^{2}\frac{\delta ^{2}}{\delta q^{2}\left(
t\right) }+m^{2}}.  \label{27}
\end{equation}%
This will be a subject of a subsequent work.

\section{\textbf{CONCLUSIONS }}

In conclusion, in the present paper we demonstrated that the quantum action
principle gives a possibility to formulate a correct quantum version of
multi-time dynamical theory with a proper probabilistic interpretation.

We thank V. A. Franke and A. V. Goltsev for useful discussions.




\end{document}